\documentclass[conference]{IEEEtran}
\usepackage[T1]{fontenc}
\usepackage{amsfonts}
\usepackage[bottom=0.6in,top=0.8in,left=0.72in,right=0.72in]{geometry}

\usepackage{graphicx}
\usepackage{caption}
\usepackage{subcaption}
\usepackage{color}

\IEEEoverridecommandlockouts
\usepackage{mathtools}
\usepackage{amsmath}
\usepackage{algorithmic,algorithm}
\usepackage{cite}
\usepackage{tabularx}

\allowdisplaybreaks

\usepackage{amsmath}
\usepackage{amsthm}

\usepackage{amsthm}

\newtheorem{theorem}{Theorem}

\newtheoremstyle{lemma}
  {0}
  {0}
  {}
  {}
  {\itshape}
  {:}
  {.5em}
  {\thmname{#1}\thmnumber{ #2}\thmnote{ (#3)}}
\theoremstyle{lemma}
\newtheorem{lemma}{Lemma}

\begin{document}
%
\title{Robot Trajectory Planning With QoS Constrained IRS-assisted Millimeter-Wave Communications} 

\author{Cristian~Tatino,~\IEEEmembership{Student Member,~IEEE,}
        Nikolaos~Pappas,~\IEEEmembership{Member,~IEEE,}
        and~Di~Yuan,~\IEEEmembership{Senior Member,~IEEE}
\thanks{This work was supported in part by ELLIIT and by the European Union's Horizon 2020 research and innovation programme under the Marie Sklodowska-Curie grant agreement No. 643002 (ACT5G) and CENIIT.

Cristian~Tatino, Nikolaos~Pappas, and Di~Yuan are with Department of Science and Technology (ITN), Link\"{o}ping University,
Sweden (Email: cristian.tatino@liu.se, nikolaos.pappas@liu.se, di.yuan@liu.se)}
}%

\maketitle

\begin{abstract}
This paper considers the joint optimization of trajectory and beamforming of a wirelessly connected robot using intelligent reflective surface (IRS)-assisted millimeter-wave (mm-wave) communications. The goal is to minimize the motion energy consumption subject to time and communication quality of service (QoS) constraints. This is a fundamental problem for industry 4.0, where robots may have to maximize their battery autonomy and communication efficiency. In such scenarios, IRSs and mm-waves can dramatically increase the spectrum efficiency of wireless communications providing high data rates and reliability for new industrial applications.

We present a solution to the optimization problem that exploits mm-wave channel characteristics to decouple beamforming and trajectory optimizations. Then, the latter is solved by a successive-convex optimization (SCO) algorithm. The algorithm takes into account the obstacles' positions and a radio map and provides solutions that avoid collisions and satisfy the QoS constraint. Moreover, we prove that the algorithm converges to a solution satisfying the Karush-Kuhn-Tucker (KKT) conditions. 

\end{abstract}




%
\IEEEpeerreviewmaketitle

\section{Introduction}
\label{sec:Intro}
Robotic and wireless technologies are playing a crucial role in Industry 4.0. leading to full automation of manufacturing processes, warehousing, and logistics~\cite{Ericsson1}. In such scenarios, an increasing number of robots and the rising of new industrial applications, such as augmented and virtual reality for assisted manufacturing, may require Gbps peak data rates~\cite{6G}. Millimeter-wave (mm-wave) communications can provide a huge amount of bandwidth for satisfying data rate requirements for industrial applications~\cite{Cheffena}. However, high blockage sensitivity reduces communication reliability while robots move in environments with obstacles. These must be avoided by the robot trajectory, which highly affects mm-wave performance as it determines whether the robot is in line-of-sight (LOS) or non-line-of-sight (NLOS). 

To enhance coverage in mm-waves scenarios, intelligent reflective surfaces (IRSs) have been proposed~\cite{IRSmmwave3}. IRSs consist of arrays of reflective elements that can be electronically controlled to adjust the angle and the phase of the reflected signals to be either added coherently or destructively to the receiver. This represents a low-cost solution to provide alternative paths to the direct link in case of blockages. However, phase shifters must be set according to the channel that depends on the robot trajectory. Moreover, robots have tasks that are usually characterized by stringent deadlines and are equipped with batteries that limit the operational time and productivity. When a robot is out of energy, its battery must be charged with a cost of time and power. Hence, in this work, we consider wirelessly connected robots in IRS-assisted mm-wave scenarios, where trajectory and beamforming are optimized to minimize the motion energy consumption and satisfy collision avoidance, time, and QoS constraints.

Energy minimization has been one of the most important problems in robot trajectory optimization~\cite{En1,CoCP}. In~\cite{En1}, the authors show up to $50$\% of energy-saving when an optimal control of the robot's speed is performed. Wirelessly connected robots are considered in~\cite{CoCP} that proposes a joint robot communication and motion energy minimization by controlling transmit power and robot's speed along a fixed trajectory. However, the potentials of mm-waves transmissions for wirelessly connected robots need to be further explored, even though some studies have been performed for unmanned aerial vehicle (UAV) scenarios~\cite{UAVmmWave1}. A joint trajectory and IRS beamforming optimization is proposed for unmanned aerial vehicle (UAV) in~\cite{UAVIRS2}. However, the authors of~\cite{UAVIRS2} consider low frequency communications and aim to maximize the average achievable rate at the users that can lead to paths with high energy consumptions. Moreover, thanks to the possibility of controlling the altitude, UAVs can avoid most of the obstacles and NLOS conditions.

\subsection{Contributions}
\label{sec:rel}
In this paper, we consider a novel robot trajectory optimization problem with quality of service (QoS) constrained IRS-assisted mm-wave communications. Specifically, the problem aims to minimize motion energy consumption while avoiding collisions with obstacles, and satisfying minimum average data rate and time constraints to reach the final position. To the best of our knowledge, energy-efficient trajectory planning problems for wirelessly connected robots have not been considered in mm-wave industrial scenarios. We consider the mutual effects of joint trajectory and beamforming optimization on the energy consumption and achieved data rate. We propose a successive convex optimization (SCO) algorithm that can find feasible trajectories by using a radio map and environment information. We prove that the algorithm converges to a point satisfying KKT conditions, and we show the results for several system parameters. Moreover, we show how IRSs, by enhancing the coverage, can increase the robot motion energy efficiency.

\section{System Model}
\label{sec:Ass}
We consider an industrial scenario, e.g., an industrial plant, where a robot moves from a starting position $q_{s}$ to its goal $q_{d}$ within a time horizon of fixed duration that is defined by the task's deadline of the robot. The robot moves on the horizontal plane of a restricted area characterized by the presence of 3D obstacles. These are represented by a set $\mathcal{O}$ of cylinders with elliptic bases and given heights.Note that an arbitrarily shaped obstacle can be approximated by the intersection and the union of several convex shapes~\cite{Borrelli}. The area is covered by an AP using mm-waves to which the robot needs to transmit uplink data by maintaining a minimum average rate requirement ($r_{min}$). The robot is equipped with a single antenna, whereas, the AP is equipped with $N$ antennas. The robot-AP communication is assisted by an IRS consisting of $M$ reflective elements. These are managed by a controller, which shares the channel state information with the AP. A scheduler, which we assume to be co-located with the controller, optimizes the robot trajectory and the beamforming to minimize the motion energy consumption. We consider both active and passive beamforming at the AP and IRS, respectively. Hereafter, we consider the following notations: $(.)^{T}$ and $(.)^{H}$, represents the transpose and the conjugate transpose, respectively; $diag(.)$ returns the diagonalization of a vector and $arg(.)$ denotes the phase of a complex number. Finally, $\lVert . \lVert_{n}$ represents the $n$-norm.
 
\begin{figure}[tb]
	\centering
	\includegraphics[width=6cm]{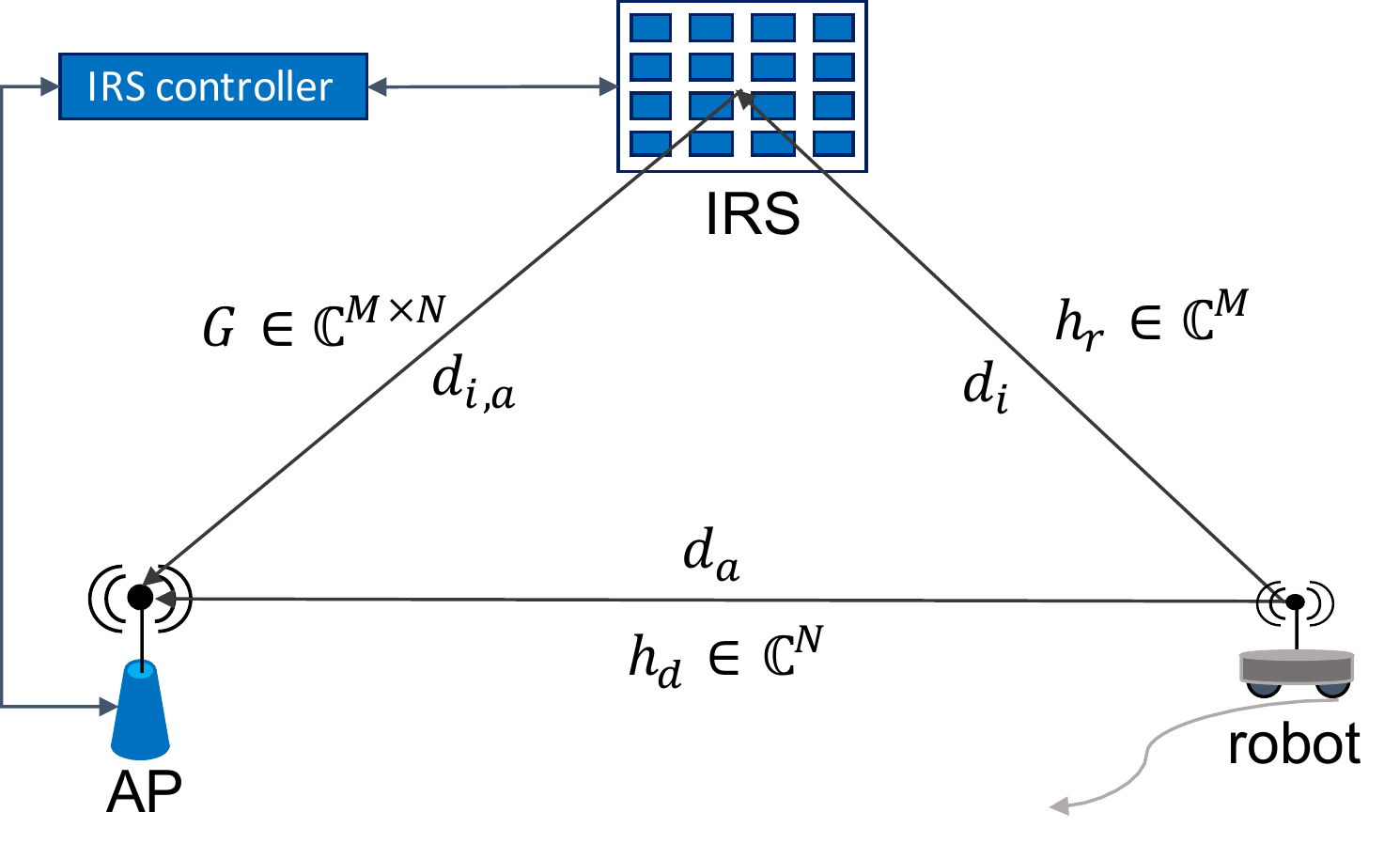}
	\caption[]{A scenario consisting of an IRS-aided robot uplink communication.\vspace{-1em}}
	\label{fig:Scen}
\end{figure}

The robot must avoid collisions with obstacles and reach the destination within a deadline that defines the time horizon. This is devided in $K$ timeslots, $k=0,...,K$, of duration $\Delta_{t}$. In each timeslot, we assume that the location of the robot and channel conditions do not change. Moreover, within a timeslot, the robot can travel for a maximum distance of $D_{max}=v_{max}\Delta_{t}$, where, $v_{max}$ is the maximum speed. A trajectory of the robot can be represented as a sequence $\boldsymbol{q}=[q_{0},q_{1},...,q_{K}]$, where $q_{0}=q_{s}$ and $q_{K}=q_{d}$. The term $q_{k}=[x_{k},y_{k}], \; k=0,...,K,$ represents the Cartesian coordinates of the robot on the horizontal plane in the $k$-th timeslot. Let $q_{a}=[x_{a},y_{a}]$ and $q_{i}=[x_{i},y_{i}]$ represents the fixed positions of the AP and the IRS, respectively. The altitude of the robot is fixed at its antenna height $z_{r}$, whereas $z_{a}$ represents the height at which the AP is installed and $z_{i}$ the height of the IRS. Let $v_{k}$ be the speed of the robot at the $k$-th timeslot, then, the motion energy consumption for DC motor-equipped robot along the path can be written as in~\cite{CoCP}: 
\begin{align*} 
E=&\sum_{k=1}^{k=K}E_{k}=\sum_{k=1}^{k=K}c_{1}v_{k}^{2}\Delta_{t}+c_{2}v_{k}\Delta_{t}+c_{3}\Delta_{t}=\\
&\sum_{k=1}^{K} c_{1} \frac{\lVert q_{k}-q_{k-1}\lVert_{2}^{2}}{\Delta_{t}}+c_{2}\lVert q_{k}-q_{k-1}\lVert_{2}+c_{3}\Delta_{t}\stepcounter{equation}\tag{\theequation}\label{eq:en_mot},
\end{align*} 
where, $E_{k}$ is the energy consumption in the $k$-th timeslot, and $c_{1}$, $c_{2}$, and $c_{3}$ are positive constants depending on the characteristics of the robot and external load.

\subsection{Channel Model}
\label{sec:Channel}
As shown in Fig.~\ref{fig:Scen}, let $h_{r}\in \mathbb{C}^{M}$ be the channel vector between the robot and the IRS and $G \in \mathbb{C}^{M\times N}$ denote the channel matrix between the IRS and the AP. The direct channel between the AP and the robot is represented by vector $h_{d}\in \mathbb{C}^{M}$. Then, the received baseband signal at the AP at timeslot $k$ can be written as follows:
\begin{align*} 
y_{k}&=\left(h_{r,k}^{H} \Phi_{k} G_{k}+h_{d,k}^{H}\right)w_{k}\sqrt{p_{t}}s_{k}+\eta_{k} \stepcounter{equation}\tag{\theequation}\label{eq:rec_sig},
\end{align*} 
where, $s_{k}$ and $p_{t}$ are the transmit signal and the transmit power in the uplink, respectively, $\eta_{k} \sim \mathcal{CN}(0,\sigma^{2})$ denotes the additive white Gaussian noise (AWGN). The term $w_{k} \in \mathbb{C}^{M}$ is the normalized beamforming vector at the AP, and $\Phi_{k}=\mbox{diag}\left(e^{j\theta_{1,k}},...,e^{j\theta_{M,k}}\right)$ is a diagonal matrix that accounts for the phase shifts $\theta_{m,k} \in  \left[ 0,2\pi  \right]$ associated with reflective elements of the IRS. Due to the high path loss of mm-wave transmissions, signals that are reflected more than once are subject to severe attenuations and are not considered in~\eqref{eq:rec_sig}. Thus, the received signal-to-noise ratio (SNR) at timeslot $k$ can be written as follows:
\begin{align*} 
\mbox{SNR}_{k}=\frac{\lvert \Big(h_{r,k}^{H} \Phi_{k} G_{k}+h_{d,k}^{H}\Big)w_{k}\rvert^{2}}{\sigma^{2}}p_{t},\stepcounter{equation}\tag{\theequation} \label{eq:snr}
\end{align*} 
where, the superscript $H$ represents the hermitian. Moreover, let $d_{i,k}=\sqrt{(z_{r}-z{i})^{2}+\lVert q_{k}- q_{i}\lVert^{2}_{2}}$, and $d_{a,k}=\sqrt{(z_{r}-z{a})^{2}+\lVert q_{k}- q_{a}\lVert^{2}_{2}}$ be the robot-IRS and robot-AP distances, respectively. Then, channel vectors $h_{r,k}$ and $h_{d,k}$ can be modeled as follows:
\vspace{-0.5em}
\noindent\begin{tabularx}{0.5\textwidth}{@{}XX@{}}
  \begin{equation}
h_{r,k}=\sqrt{\rho d_{i,k}^{-\nu}}\tilde{h}_{r,k},\label{eqn:cha_r}
  \end{equation} &
  \begin{equation}
h_{d,k}=\sqrt{\rho d_{a,k}^{-\mu}}\tilde{h}_{d,k}, \label{eqn:cha_d}
  \end{equation} 
\end{tabularx}
where, $\tilde{h}_{r,k} \sim \mathcal{CN}(0,I)$ and $\tilde{h}_{d,k} \sim \mathcal{CN}(0,I)$ are complex gaussian vectors whose elements are independent and identical distributed (i.i.d) with zero means and unit variances. The term $\rho$ is the path loss at the reference distance of $1$ m, and $\nu$ and $\mu$ are the path loss exponents of the reflected and direct channels, respectively. 

Finally, let $\boldsymbol{\mbox{r}}=\Big[\mbox{r}_{0},\mbox{r}_{1},...,\mbox{r}_{K}\Big]$ be a vector, the elements of which represent the rates at timeslots $k=0,1,...,K$. Thus, by using the Shannon's formula, the average rate ($\bar{\boldsymbol{\mbox{r}}}$) for a trajectory $\boldsymbol{q}$ is given by:
\begin{align*} 
&\bar{\boldsymbol{\mbox{r}}}=\frac{1}{K}\sum_{k=0}^{K}\mbox{r}_{k}=\frac{\mbox{B}_{w} }{K}\sum_{k=0}^{K}\mbox{log}_{2}\left(1+SNR_{k}\right)=\frac{\mbox{B}_{w} }{K}\times\stepcounter{equation}\tag{\theequation} \label{eq:avg_rate}\\
&\sum_{k=0}^{K} \mbox{log}_{2}\left(1+\frac{\lvert \Big(\sqrt{\rho d_{i,k}^{-\nu}}\tilde{h}_{r,k}^{H} \Phi_{k} G_{k}+\sqrt{\rho d_{a,k}^{-\mu}}\tilde{h}_{d,k}^{H}\Big)w_{k}\rvert^{2}}{\sigma^{2}}p_{t}\right),
\end{align*} 
where, \eqref{eqn:cha_r} and~\eqref{eqn:cha_d} are used in~\eqref{eq:snr} and $\mbox{B}_{w}$ represents the system bandwidth. We can observe that $\bar{\boldsymbol{\mbox{r}}}$ is a function of $\Phi_{k}$, $w_{k}$, and $q_{k}$. The latter is included in $d_{i,k}$, and $d_{a,k}$. Thus, the robot's position and beamforming affect the rate, which in turn affects the trajectory. In the next section, we formulate the problem of minimizing motion energy consumption, maintaining a certain minimum average data rate considering both the beamforming and trajectory optimization.


\section{Problem Formulation}
\label{sec:Prob}
In this section, we formulate the problem introduced in 
Section~\ref{sec:Ass}. Let $\boldsymbol{\Phi}=[\Phi_{0},\Phi_{1},...,\Phi_{K}]$ and $\boldsymbol{w}=[w_{0},w_{1},...,w_{K}]$, then the joint robot trajectory and beamforming problem can be formulated as follows:
\begin{subequations}
\begin{align}
        P1:&\min_{\boldsymbol{q},\boldsymbol{\Phi},\boldsymbol{w}}E\label{opt}\\ 
        \text{s.t.}& \;\bar{\boldsymbol{\mbox{r}}}\ge \mbox{r}_{min}\label{con_Rate},\\
        		&\lVert q_{k}-q_{k-1} \lVert_{2}  \le D_{max}, \; \; k=1,...,K, \label{Con_Dist}\\
		&q_{0}=q_{s}, \; \; q_{K}=q_{d},\label{Con_Traj}\\
		&\left(q_{k}-q_{c,o}\right)^{T}P_{o}^{-1}\left(q_{k}-q_{c,o}\right)\ge d_{s},  \; \forall k, \forall o\in \mathcal{O}, \label{Con_Ob}\\
		&\lVert w_{k} \lVert_{2}^{2}  \le 1, \; \; \forall k, \label{Con_w}\\
		&\Phi_{k}=\mbox{diag}\left(e^{j\theta_{1k}},...,e^{j\theta_{Mk}}\right), \; \forall k, \label{Con_Phi}\\
		&0 \le \theta_{m,k}\le 2\pi, \; \; \forall m, \; \forall k, \label{Con_theta}
\end{align}
\end{subequations}
where, the objective function~\eqref{opt} represents the total robot motion energy consumption along the trajectory given by~\eqref{eq:en_mot}. The first constraint~\eqref{con_Rate} represents the QoS requirement to complete the task, where $\bar{\boldsymbol{\mbox{r}}}$ is defined in~\eqref{eq:avg_rate} and $\mbox{r}_{min}$ is the minimum required average rate. Constraints~\eqref{Con_Dist} allow the robot to move in a timeslot for a maximum distance of $D_{max}$, whereas~\eqref{Con_Traj} fix the starting and the goal positions. To avoid collisions with obstacles, we include~\eqref{Con_Ob}. As described in Section~\ref{sec:Ass}, each obstacle $o \in \mathcal{O}$ is characterized by an ellipsoid shape on the horizontal plane, with a center $q_{c,o}$, and a symmetric and positive definite matrix $P_{o}$. The latter defines the length of the axis and the rotation of the ellipse, and $d_{s}\ge1$ represents a safety distance between the robot and the obstacle. Constraints~\eqref{Con_w} and~\eqref{Con_theta} impose the norm of $w_{k}$ to be at most one and $\theta_{m,k}$ to be continuous, respectively.
 
Problem $P1$ is non-linear and non-convex. However, we show that, for each robot trajectory $\boldsymbol{q}$, it is possible to find closed-forms of $\boldsymbol{\Phi}$ and $\boldsymbol{w}$ that maximize the average rate. More precisely, $\Phi_{k}$ and $w_{k}$ are not contributing to the cost in~\eqref{opt}. Thus, when the right-hand side (RHS) of~\eqref{con_Rate} is maximized given a certain trajectory, we obtain a problem with a larger feasible region, of which the optimum is equivalent to that of $P1$. This equivalent problem of constrained trajectory optimization is solved in Section~\ref{sec:Traj_Opt} by using an SCO algorithm.




%

\subsection{Average Rate Maximization}
\label{sec:Avg_Rate}
In this section, we first find closed-form solutions of $\boldsymbol{\Phi}$ and $\boldsymbol{w}$ that maximize the average rate $\bar{\boldsymbol{\mbox{r}}}$ for a fixed trajectory by solving the following problem:
\begin{subequations}
\begin{align}
        P2:&\max_{\boldsymbol{\Phi},\boldsymbol{w}}\bar{\boldsymbol{\mbox{r}}}\label{opt2}\\ 
		 \text{s.t.}& \; \eqref{Con_w},\eqref{Con_Phi},\eqref{Con_theta},\nonumber
\end{align}
\end{subequations}
where, $\bar{\boldsymbol{\mbox{r}}}$ is given by~\eqref{eq:avg_rate}. We can assume that the IRS and the AP are installed with LOS between them. Since in mm-wave communications the LOS path presents a much higher gain than the sum of NLOS paths, the IRS-AP channel at timeslot $k$ can be approximated by a rank-one matrix~\cite{IRSmmwave3}: $G_{k}=\sqrt{NM}\gamma \tilde{a}_{k}\tilde{b}_{k}^{T}$, where, $\gamma=\sqrt{\rho d_{ia}^{-2}}$, and $d_{ia}$ is the AP-IRS distance that is fixed and does not depend on $q_{k}$. The path loss exponent that is associated with the LOS path between the AP and the IRS is two. The terms $\tilde{a}_{k} \in \mathbb{C}^{M}$ and $\tilde{b}_{k} \in \mathbb{C}^{N}$ are the normalized array response vectors in timeslot $k$ associated with the IRS and the AP, respectively.

Maximizing P2 is equivalent to maximizing the received SNR in each timeslot $k$~\eqref{eq:snr}. Assuming $\Phi_{k}=e^{\alpha_{k}}\widehat{\Phi}_{k}$, this problem has a closed-form solution~\cite{IRSmmwave3}:
\begin{align*} 
\alpha_{k}^{*}=-\mbox{arg}\left(\left(\tilde{b}_{k}^{T}\right)^{H}\tilde{h}_{d,k}\right),\stepcounter{equation}\tag{\theequation} \label{eq:sol_alpha}
\end{align*} 
\begin{align*} 
\widehat{\Phi}_{k}^{*}=\mbox{diag}\left(e^{-j\mbox{arg}(g_{1,k})},...,e^{-j\mbox{arg}(g_{M,k})}\right),\stepcounter{equation}\tag{\theequation}\label{eq:sol_phi}
\end{align*} 
\begin{align*} 
w_{k}^{*}=\frac{\left(e^{\alpha_{k}^{*}}\sqrt{\rho d_{i,k}^{-\nu}}\tilde{h}_{r,k}^{H} \bar{\Phi}_{k}^{*} G_{k}+\sqrt{\rho d_{a,k}^{-\mu}}\tilde{h}_{d,k}^{H}\right)^{H}}{\lVert e^{\alpha_{k}^{*}}\sqrt{\rho d_{i,k}^{-\nu}}\tilde{h}_{r,k}^{H} \bar{\Phi}_{k}^{*} G_{k}+\sqrt{\rho d_{a,k}^{-\mu}}\tilde{h}_{d,k}^{H}\lVert_{2}},\stepcounter{equation}\tag{\theequation}\label{eq:sol_w}
\end{align*} 
where, $g_{k}=\sqrt{\rho d_{iak}^{-2}}\left(\tilde{h}_{r,k}^{*}\circ\tilde{a}_{k}\right)$ and $\left(\circ\right)$ denotes the elementwise product. By putting~\eqref{eq:sol_alpha}, \eqref{eq:sol_phi}, and~\eqref{eq:sol_w} into~\eqref{eq:snr}, we obtain the following optimal SNR expression for timeslot $k$:
\begin{align*} 
\mbox{SNR}^{*}_{k}&=\Big(N\lvert \rho \rvert \lvert \gamma \rvert^{2} \lVert \tilde{h}_{r,k}\lVert_{1}^{2}d_{i,k}^{-\nu}+\\
2\sqrt{N} \lvert \rho &\rvert \gamma \lVert \tilde{h}_{r,k}^{H}\lVert_{1} \lvert \tilde{b}_{k}^{T}\tilde{h}_{d,k} \rvert d_{i,k}^{-\nu/2}d_{a,k}^{-\mu/2}+\rho \lVert \tilde{h}_{d,k}\lVert_{2}^{2}d_{a,k}^{-\mu} \Big) \frac{p_{t}}{\sigma^{2}}=\\
&\Big(A d_{i,k}^{-\nu}+ B  d_{i,k}^{-\nu/2} d_{a,k}^{-\mu/2}+ Cd_{a,k}^{-\mu} \Big) \frac{p_{t}}{\sigma^{2}},\stepcounter{equation}\tag{\theequation} \label{eq:snr_max}
\end{align*} 
where, in the last equality, we have highlighted the dependence of $\mbox{SNR}^{*}_{k}$ on the robot position $q_{k}$ through the terms $d_{a,k}$ and $d_{i,k}$. However, $\nu$ and $\mu$ depend on the scattering environment. For this reason, starting from~\eqref{eq:snr_max}, we estimate $\nu$, $\mu$, and other parameters, i.e., $A$, $B$, and $C$,  by fitting~\eqref{eq:snr_max} with a radio map. The radio map provides the averaged optimal $\mbox{SNR}^{*}$ for each position that is obtained by computing \eqref{eq:sol_alpha}, \eqref{eq:sol_phi}, and~\eqref{eq:sol_w} from a set of channel measurements. This procedure results in:
\begin{align*} 
\widehat{\mbox{SNR}}^{*}_{k}&=\Big(\widehat{A} d_{i,k}^{-\widehat{\nu}}+ \widehat{B}  d_{i,k}^{-\widehat{\nu}/2} d_{a,k}^{-\widehat{\mu}/2}+ \widehat{C}d_{a,k}^{-\widehat{\mu}} \Big) \frac{p_{t}}{\sigma^{2}},\stepcounter{equation}\tag{\theequation} \label{eq:snr_max_est}
\end{align*} 
where, $\widehat{A}$, $\widehat{B}$, $\widehat{C}$, $\widehat{\nu}$, and $\widehat{\mu}$ are the estimated parameters.

We can use~\eqref{eq:snr_max_est} in~\eqref{eq:avg_rate} to obtain an estimation of the maximum rate resulting from the beamforming optimization:
\begin{align*} 
&\bar{\boldsymbol{\mbox{r}}}^{*}=\frac{1}{K}\sum_{k=0}^{K}\mbox{r}^{*}_{k}=\frac{\mbox{B}_{w}}{K}\sum_{k=0}^{K}\mbox{log}_{2}\left(1+\widehat{\mbox{SNR}}^{*}_{k}\right)=\stepcounter{equation}\tag{\theequation} \label{eq:avg_rate2}\\
&\frac{\mbox{B}_{w}}{K} \sum_{k=0}^{K}\mbox{log}_{2}\left(1+\Big(\widehat{A} d_{i,k}^{-\widehat{\nu}}+ \widehat{B}  d_{i,k}^{-\widehat{\nu}/2} d_{a,k}^{-\widehat{\mu}/2}+ \widehat{C} d_{a,k}^{-\widehat{\mu}} \Big) \frac{p_{t}}{\sigma^{2}}\right),
\end{align*} 
where, $\mbox{r}^{*}_{k}$ is the optimized rate at timeslot $k$. Finally, by replacing the RHS of~\eqref{con_Rate} with~\eqref{eq:avg_rate2}, we can decouple the beamforming and the trajectory optimization obtaining the following problem:
\vspace{-0.5em}
\begin{subequations}
\begin{align}
        P3:&\min_{\boldsymbol{q},\boldsymbol{\Phi},\boldsymbol{w}}E\label{opt3}\\
        \text{s.t.}& \;\bar{\boldsymbol{\mbox{r}}}^{*}\ge \mbox{r}_{min}\label{con_Rate3},\\
        		&~\eqref{Con_Dist},~\eqref{Con_Traj},~\eqref{Con_Ob},\nonumber
\end{align}
\end{subequations}
where, $E$ in~\eqref{opt3} is the robot energy consumption that is given by~\eqref{eq:en_mot}. This problem is equivalent to P1, but it considers the estimated maximum rate, $\bar{\boldsymbol{\mbox{r}}}^{*}$. 

\section{Trajectory Optimization}
\label{sec:Traj_Opt}
In this section, we provide an algorithm to solve trajectory optimization problem $P3$. We first derive the following lemma:
\begin{lemma}\label{lemma1}
Given $c_{1}\ge0$, $c_{2}\ge0$, and $c_{3}\ge0$, the objective function of $P3$~\eqref{eq:en_mot} is a convex function of $\boldsymbol{q}$.
\end{lemma}
\vspace{-0.5em}
\begin{proof}
We prove Lemma 1 by induction. Let $E|_{K=n}$ be the energy consumption along the path~\eqref{eq:en_mot} when $K=n$: $\sum_{k=1}^{n} c_{1} \frac{\lVert q_{k}-q_{k-1}\lVert_{2}^{2}}{\Delta_{t}}+c_{2}\lVert q_{k}-q_{k-1}\lVert_{2}+c_{3}\Delta_{t}$. We first prove that $E|_{K=1}$ is convex and then, by assuming that convexity holds for $E|_{K=n-1}$ we prove that $E|_{K=n}$ is a convex function of $\boldsymbol{q}=[q_{0},...,q_{n}]$. It is easy to show that $E|_{K=1}=c_{1} \frac{\lVert q_{1}-q_{0}\lVert_{2}^{2}}{\Delta_{t}}+c_{2}\lVert q_{1}-q_{0}\lVert_{2}+c_{3}\Delta_{t}$ is a convex function of $q_{0}$ and $q_{1}$ because it consists of the sum of two convex functions, i.e., $c_{1} \frac{\lVert q_{1}-q_{0}\lVert_{2}^{2}}{\Delta_{t}}$ and $c_{2}\lVert q_{1}-q_{0}\lVert_{2}$, and a constant term. Assume that $E|_{K=n-1}$ is convex, then we have that $E|_{K=n}=E|_{K=n-1}+c_{1} \frac{\lVert q_{n}-q_{n-1}\lVert_{2}^{2}}{\Delta_{t}}+c_{2}\lVert q_{n}-q_{n-1}\lVert_{2}+c_{3}\Delta_{t}$. By following the same reasoning, we can observe that $E|_{K=n}$ is the sum of three convex functions of $\boldsymbol{q}=[q_{0},...,q_{n}]$: $E|_{K=n-1}$ that is convex by hypothesis, $\frac{\lVert q_{n}-q_{n-1}\lVert_{2}^{2}}{\Delta_{t}}$, and $c_{2}\lVert q_{n}-q_{n-1}\lVert_{2}$.
\end{proof}\vspace{-0.5em}
However, $P3$ is not convex because the RHS of~\eqref{con_Rate3} and~\eqref{Con_Ob} are not concave functions of $q_{k}$. For this reason, we perform a convex local approximation of these two constraints and solve the problem iteratively by using an SCO algorithm. Starting from constraint~\eqref{con_Rate3}, we have the following:
\begin{lemma}\label{lemma2}
Given $\bar{A}\ge0$, $\bar{B}\ge0$, $\bar{C}\ge0$, $\bar{\nu}\ge0$, and $\bar{\mu}\ge0$, $\bar{\boldsymbol{\mbox{r}}}^{*}$ is a convex function of $d_{a,k}$ and $d_{i,k}$ with $k=0,...,K$.
\end{lemma}
\vspace{-0.5em}
\begin{proof}
See Appendix A.
\end{proof}
\vspace{-0.5em}

Thus, since any convex function can be lower-bounded by its first-order Taylor expansion, we can write the following:
\vspace{-0.5em}
\begin{align*} 
&\bar{\boldsymbol{\mbox{r}}}^{*}\ge\bar{\boldsymbol{\mbox{r}}}_{app}^{*}=\frac{1}{K}\sum_{k=0}^{K}\mbox{r}_{app,k}^{*}=\\
&\frac{\mbox{B}_{w}}{K}\sum_{k=0}^{K} \mbox{log}_{2}\left(1+\Big(\bar{A} d_{i,0,k}^{-\bar{\nu}}+ \bar{B}  d_{i,0,k}^{-\bar{\nu}/2} d_{a,0,k}^{-\bar{\mu}/2}+\bar{C} d_{a,0,k}^{-\bar{\mu}} \Big) \frac{p_{t}}{\sigma^{2}}\right)+\\
& \nabla \bar{\boldsymbol{\mbox{r}}}^{*}|^{T}_{\big(d_{a,0,k},d_{i,0,k}\big)}
\begin{bmatrix}
d_{a,k}-d_{a,0,k}\\ 
d_{i,k}-d_{i,0,k}
\end{bmatrix},\stepcounter{equation}\tag{\theequation} \label{eq:avg_rate_appx}
\end{align*}
where, $d_{a,0,k}$ and $d_{i,0,k}$ are the expansion points. The gradient ($\nabla$) with respect to  $d_{a,k}$ and $d_{i,k}$ is given by:
\vspace{-0.5em}
\begin{align*} 
\nabla \bar{\boldsymbol{\mbox{r}}}^{*}=\frac{\mbox{B}_{w}}{K}\sum_{k=0}^{K}
\begin{bmatrix}
\frac{\Big(-\bar{\nu}\bar{A} d_{i,k}^{-\bar{\nu}-1}-\bar{\nu}/2\bar{B}  d_{i,k}^{-\bar{\nu}/2-1} d_{a,k}^{-\bar{\mu}/2}\Big)\frac{p_{t}}{\sigma^{2}}}{\mbox{ln}2\Bigg(1+\Big(\bar{A} d_{i,k}^{-\bar{\nu}}+ \bar{B}  d_{i,k}^{-\bar{\nu}/2} d_{a,k}^{-\bar{\mu}/2}+ \bar{C} d_{a,k}^{-\bar{\mu}} \Big)\frac{p_{t}}{\sigma^{2}}\Bigg)}\\ \\
\frac{\Big(-\bar{\mu}\bar{C} d_{a,k}^{-\bar{\mu}-1}-\bar{\mu}/2\bar{B}  d_{i,k}^{-\bar{\nu}/2} d_{a,k}^{-\bar{\mu}/2-1}\Big)\frac{p_{t}}{\sigma^{2}}}{\mbox{ln}2\Bigg(1+\Big(\bar{A} d_{i,k}^{-\bar{\nu}}+ \bar{B}  d_{i,k}^{-\bar{\nu}/2} d_{a,k}^{-\bar{\mu}/2}+ \bar{C} d_{a,k}^{-\bar{\mu}} \Big)\frac{p_{t}}{\sigma^{2}}\Bigg)}
\end{bmatrix}.\stepcounter{equation}\tag{\theequation} \label{eq:grad_rate}
\end{align*}
We now consider the following lemma:
\begin{lemma}\label{lemma3}
Given $d_{i,k}=\lVert q_{k}- q_{i}\lVert_{2}$ and $d_{a,k}=\lVert q_{k}- q_{a}\lVert_{2}$ and non-negative parameters $\bar{A}$, $\bar{B}$, $\bar{C}$, $\bar{\nu}$, and $\bar{\mu}$, $\bar{\boldsymbol{\mbox{r}}}_{app}^{*}$ is a concave function of $q_{k}$.
\end{lemma}
\vspace{-0.5em}
\begin{proof}
See Appendix B.
\end{proof}

Thus, in a small neighborhood of $d_{a,0,k}$ and $d_{i,0,k}$, we can approximate $\bar{\boldsymbol{\mbox{r}}}^{*}$ with a concave function of $q_{k}$ given by $\bar{\boldsymbol{\mbox{r}}}_{app}^{*}$. The same reasoning can be applied to~\eqref{Con_Ob} leading to the following inequality:
\begin{align*} 
&\left(q_{k}-q_{c,o}\right)^{T}P_{o}^{-1}\left(q_{k}-q_{c,o}\right)\ge\\
&\left(q_{0,k}-q_{c,o}\right)^{T}P_{o}^{-1}\left(q_{0,k}-q_{c,o}\right)+\\
&\left(q_{0,k}-q_{c,o}\right)^{T}P_{o}^{-1}\left(q_{k}-q_{0,k}\right),\stepcounter{equation}\tag{\theequation} \label{eq:grad_obs}
\end{align*}
where, the RHS is the first-order Taylor expansion of~\eqref{Con_Ob} with respect to $q_{k}$ at local point $q_{0,k}$. This is an affine function of $q_{k}$. Finally, by considering~\eqref{eq:avg_rate_appx},~\eqref{eq:grad_obs} and a starting from a feasible initial trajectory $\boldsymbol{q}_{0}$, we can solve a sequence of local convex approximations of $P3$, as shown in Algorithm 1. At each iteration $j$, Algorithm 1 solves the following problem:
\begin{subequations}
\begin{align}
        P4:&\min_{\boldsymbol{q}_{j}}\sum_{k=1}^{K} c_{1} \frac{\lVert q_{j,k}-q_{j,k-1}\lVert_{2}^{2}}{\Delta_{t}}+c_{2}\lVert q_{j,k}-q_{j,k-1}\lVert_{2}+c_{3}\Delta_{t}\label{opt3}\\ 
        \text{s.t.:}& \;\bar{\boldsymbol{\mbox{r}}}_{app,j}^{*}\ge \mbox{r}_{min}\label{con_Rate4},\\
        		&\lVert q_{j,k}-q_{j,k-1} \lVert_{2}  \le D_{max}, \; \;\forall k, \label{Con_Dist4}\\
		&q_{0}=q_{s}, \; \; q_{k}=q_{d},\label{Con_Traj4}\\
        		&\lVert q_{j,k}-q_{j-1,k} \lVert_{2}  \le T, \; \;\forall k, \label{Con_Trust}\\
		&\left(q_{j-1,k}-q_{c,o}\right)^{T}P_{o}^{-1}\left(q_{j-1,k}-q_{c,o}\right)+\nonumber\\
&\left(q_{j-1,k}-q_{c,o}\right)^{T}P_{o}^{-1}\left(q_{j,k}-q_{j-1,k}\right)\ge d_{s}, \forall k,o, \label{Con_Ob4}
\end{align}
\end{subequations}
where, $\boldsymbol{q}_{j}=[q_{j,0},q_{j,1},...,q_{j,K}]$ and $\boldsymbol{q}_{j-1}=[q_{j-1,0},q_{j-1,1},...,q_{j-1,K}]$ are the solutions of $P4$ at iteration $j$ and $j-1$, respectively. More precisely, $\boldsymbol{q}_{j-1}$ represents the local point at which the approximations at iteration $j$ of constraints~\eqref{con_Rate3} and~\eqref{Con_Ob} are computed. These approximations are valid in a trust region of $\boldsymbol{q}_{j-1}$ ($T$) that is defined by constraint~\eqref{Con_Trust}. Note that the expansion points of $\bar{\boldsymbol{\mbox{r}}}^{*}$ can be obtained from $q_{j-1,k}$, as 
$d_{i,k}=\sqrt{(z_{r}-z{i})^{2}+\lVert  q_{j-1,k}-q_{i}\lVert^{2}_{2}}$, and $d_{a,k}=\sqrt{(z_{r}-z{a})^{2}+\lVert q_{j-1,k}-q_{a}\lVert^{2}_{2}}$. Problem $P4$ is convex and it can be solved quickly by interior-point methods. Algorithm 1 stops if the sequence of solutions converges or when a maximum number of iterations ($N_{it}$) is reached. Specifically, we can prove the following:
\vspace{-0.5em}
\begin{theorem}\label{theo2}
Algorithm~1 provides a sequence of solutions that is non-increasing and converges to a KKT point of $P3$.
\end{theorem}
\vspace{-1em}
\begin{proof}
The convergence of Algorithm 1 to a KKT point of $P3$ follows from~\cite{Conv}. Moreover, the sequence of solutions provided by Algorithm 1 is non increasing because, the solution of $P4$ at iteration $j-1$, $\boldsymbol{q}_{j-1}$, is a feasible solution of minimization problem $P4$ at iteration $j$.
\end{proof}
\vspace{-0.5em}

\begin{algorithm}
\algsetup{linenosize=\tiny}
\caption{\textit{QoS Constrained Robot Trajectory Planning}}
\label{alg:Algo}
\begin{algorithmic}[1]
  \scriptsize
  \item[\textbf{\textit{Initial solution}}:]
  \STATE{j=0}
  \STATE{find an initial feasible solution $\boldsymbol{q}_{j}$}
  \STATE{Compute the motion energy consumption $E_{j}$ corresponding to $\boldsymbol{q}_{j}$ as in~\eqref{opt3}}  
  \item[\textbf{\textit{SCO}}:]
  \REPEAT {
  	\STATE{$j=j+1$}  
 	 \STATE{Obtain $\boldsymbol{q}_{j}$ and $E_{j}$ by solving $P4$ with local points $\boldsymbol{q}_{j-1}$}	 
	 } \UNTIL{$\frac{E_{j}-E_{j-1}}{E_{j-1}}\le \epsilon$ or $j\ge N_{it}$}
\end{algorithmic}
\end{algorithm}
\vspace{-0.5em}
We consider two initial solutions for Algorithm 1 that are obtained by computing the shortest path on a time expanded graph as done in~\cite{MAPP}. The edges and vertices of the graph are defined on a discrete set of positions that are free from obstacles. Then, by using the radio map, the costs of the edges are set to generate a first solution that minimizes the motion energy consumption (ME) and a second solution that maximizes the rate (MR). Algorithm 1 uses ME if this is feasible and MR, otherwise. If also the latter is not feasible, the algorithm declares infeasibility.

\begin{figure}[tb]
	\centering
	\includegraphics[width=7cm]{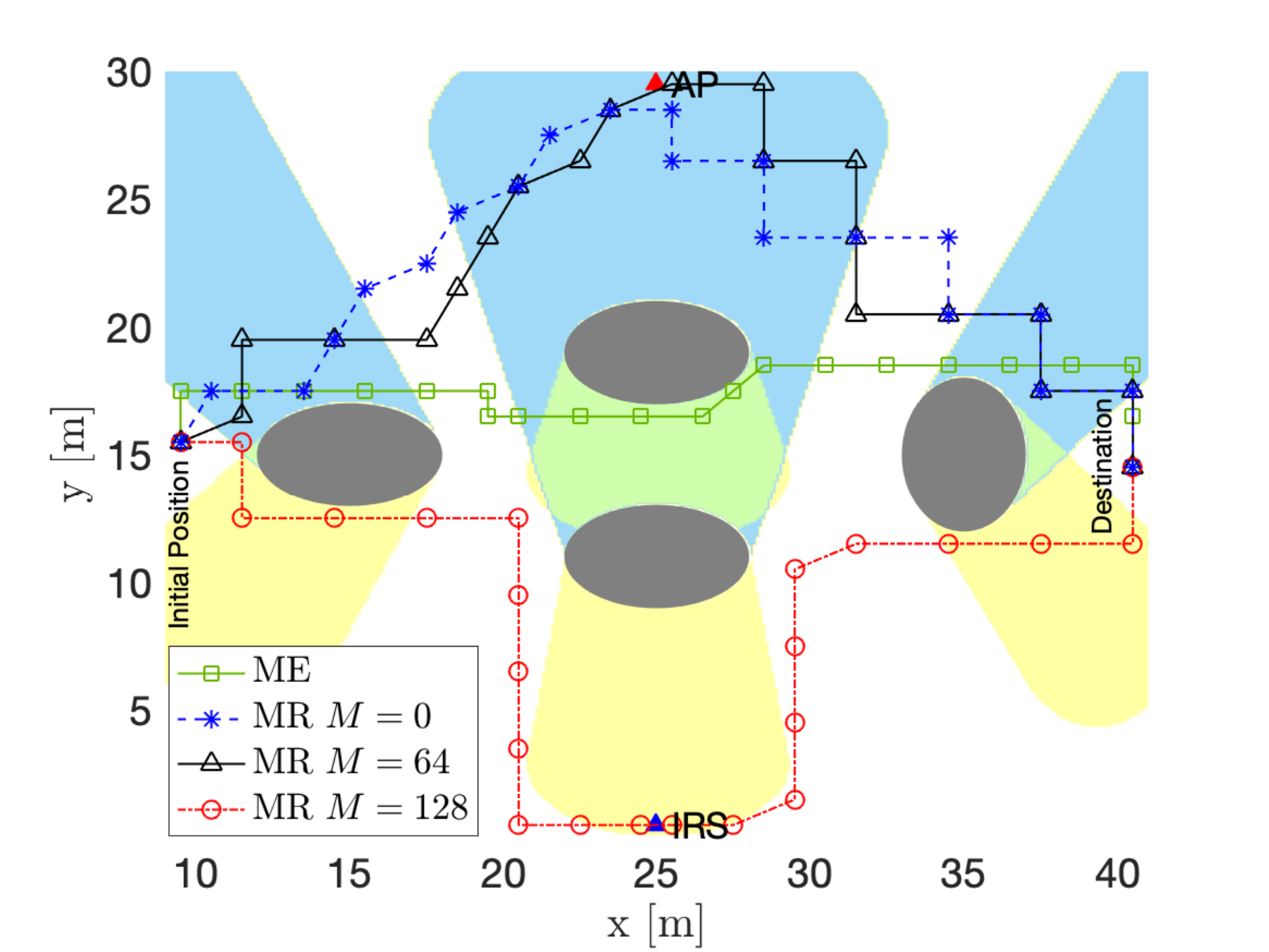}
	\caption[]{Minimum energy (ME) and maximum rate (MR) initial solutions for $K=30$ and several values of $M$. Yellow, blue, and green shaded positions are in NLOS with respect to the AP, the IRS, and both, respectively. The positions in the white area are in LOS with respect to both the IRS and the AP.\vspace{-1em}}
	\label{fig:Init}
\end{figure}
\begin{figure}[tb]
	\centering
	\includegraphics[width=7cm]{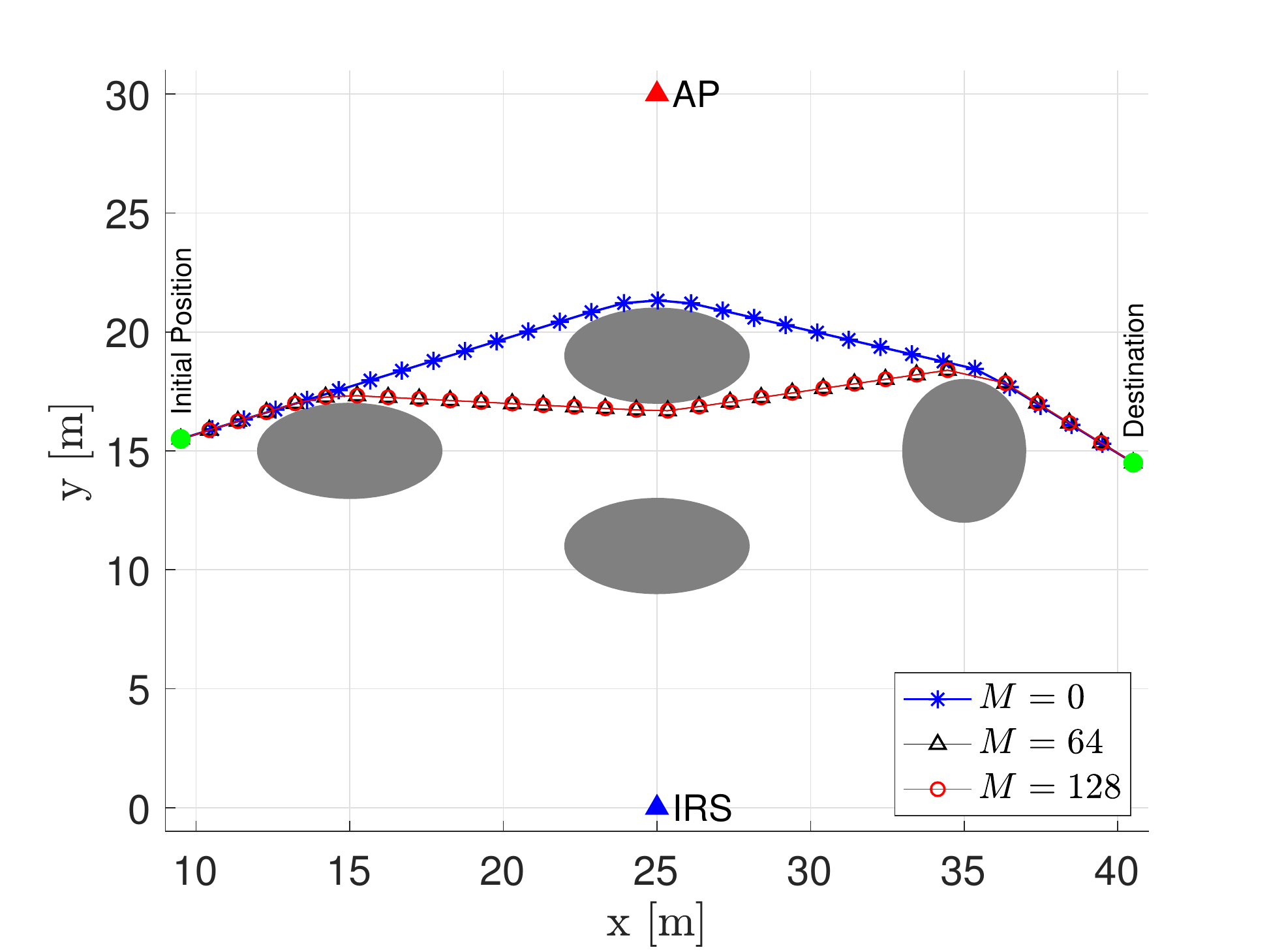}
	\caption[]{Robot trajectories to which Algorithm 1 converges for $r_{min}=2$\,Gbps and several values of $M$.\vspace{-1em}}
	\label{fig:Path2}
\end{figure}

\section{Numerical Results}
\label{sec:Res}
In this section, we provide a numerical validation of Algorithm 1 for solving $P3$. For our simulations, we consider a $50 \times 30$\,m$^{2}$ rectangular-shaped indoor scenario. The robot's starting position is $[9.5,15.5]$, whereas the destination is $[40.5,14.5]$. Ellipse obstacles, with length, width, and height of $6$\,m, $4$\,m, and $2$\,m, respectively, are placed as in Fig.~\ref{fig:Init}, where they are represented by grey shaded areas. In this scenario, there is an AP and an IRS placed at [25,30] and [25,0], respectively, operating in the $60$~GHz band as in~\cite{MMInd2}, with bandwidth $B_{w}=200$ MHz. The height of the AP is $5$\,m, whereas, we set the heights of the IRS and the robot's antenna equal to $2.5$ and $0.5$\,m, respectively. This scenario includes several robot-AP and robot-IRS channel conditions, i.e., LOS and NLOS positions. NLOS positions with respect to the AP, the IRS, and both are represented in Fig.~\ref{fig:Init} by yellow, blue, and green shaded areas, respectively. 

In Fig.~\ref{fig:Init}, we also show ME and MR initial solutions for several values of $M$. Note that MR initial solutions consider trajectories that avoid NLOS areas with respect to either the AP or to the IRS according to the number of reflective elements of the latter. Similar to~\cite{MMInd2}, the path loss at a reference distance of 1 meter is $68$ dB, and the path loss exponent of the robot-AP and robot-IRS channels are set to $2$, for LOS cases, and $4.5$, for positions in NLOS. The path loss exponent of the IRS-AP channel is set to $2$. Moreover, we set the transmit and the noise powers to $20$\,dBm, and $-80$\,dBm, respectively. We show the results, for several values of $M$ and $r_{min}$, and we set the following parameters as follows: $N=16$, $K=30$, $\Delta_{t}=1$\,s, $v_{max}=3$\,m/s, $d_{s}=1.35$, $N_{it}=100$, $\epsilon=0.01$, $T=1$\,m, $c_{1}=4.39$, $c_{2}=24.67$, and $c_{3}=14.77$~\cite{CoCP}. Finally, we can obtain~\eqref{eq:snr_max_est} by estimating $\widehat{A}\ge0$, $\widehat{B}\ge0$, $\widehat{C}\ge0$, $\widehat{\nu}\ge0$, and $\widehat{\mu}\ge0$ by fitting~\eqref{eq:snr_max}  on a radio map by solving a non-linear least squares problem. The radio map is obtained from the average of $10.000$ channel measurements on a grid of $500\times 300$ points. 

In Fig.~\ref{fig:Path2}, we show robot trajectories resulting from Algorithm 1 for $r_{min}=2$\,Gbps and several values of $M$. For $M=0$, we can observe that the robot avoids NLOS areas with respect to the AP, and Algorithm 1 uses initial solution MR. When $M$ increases, the IRS enhances the coverage and the robot can find a trajectory with lower energy consumption by using initial solution ME. The resulting trajectory crosses the NLOS area with respect to both the AP and the IRS. Note that, for $r_{min}=2$\,Gbps, values of $M$ that are higher than $64$ do not provide further gain, and the trajectories for $M=64$ and $M=128$ coincide. 

\begin{figure}[tb]
	\centering
	\includegraphics[width=7cm]{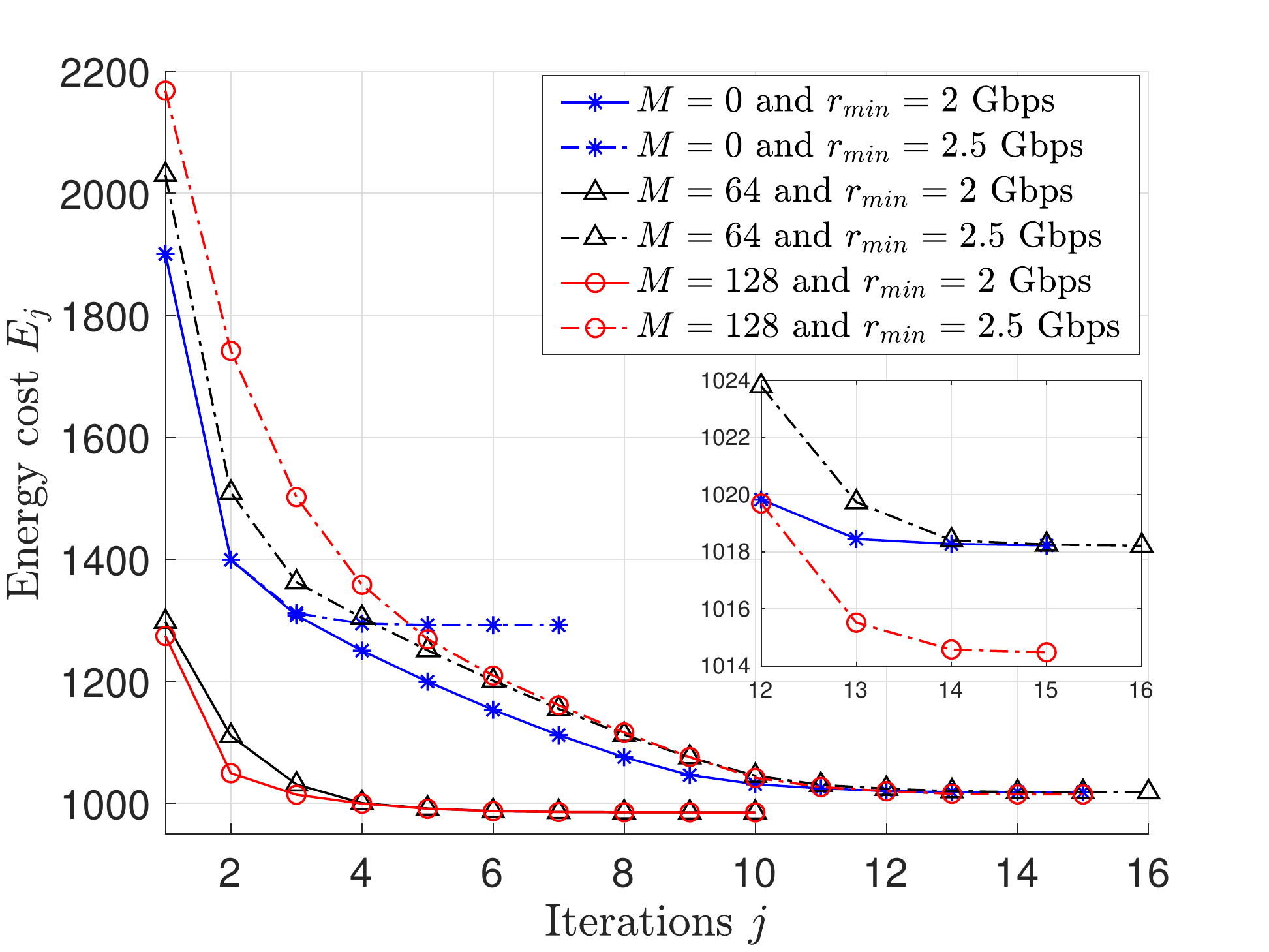}
	\caption[]{Energy consumption corresponding to the sequence of solutions $\boldsymbol{q}_{j}$ provided by Algorithm 1 for several values of $M$ and $r_{min}$.\vspace{-1em}}
	\label{fig:En_it}
\end{figure}

\begin{figure}[tb]
	\centering
	\includegraphics[width=7cm]{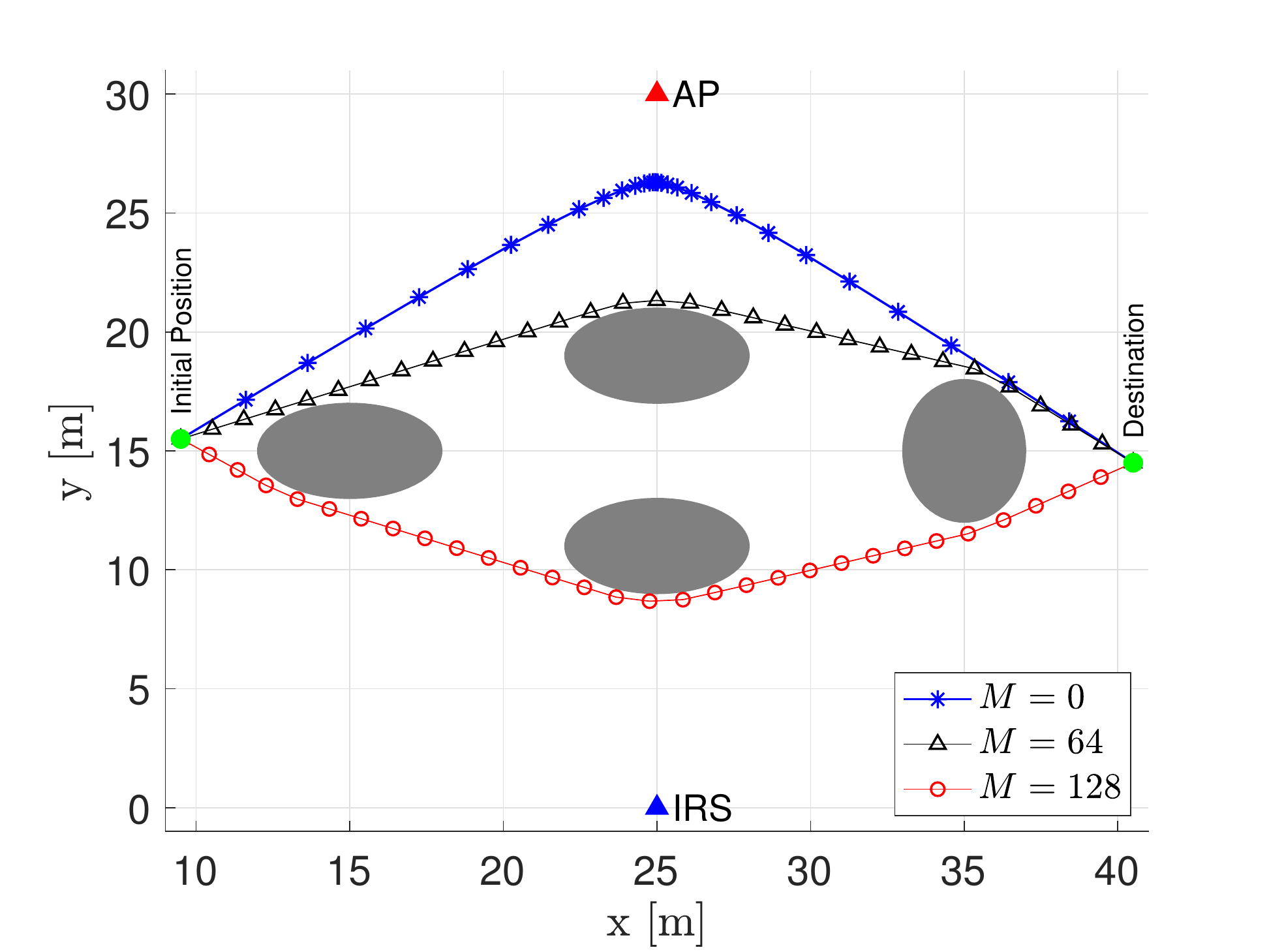}
	\caption[]{Robot trajectories to which Algorithm 1 converges for $r_{min}=2.5$\,Gbps and several values of $M$.\vspace{-1em}}
	\label{fig:Path2_5}
\end{figure}

This can be better observed in Fig.~\ref{fig:En_it}, where we show the energy consumption corresponding to the sequence of solutions $\boldsymbol{q}_{j}$ provided by Algorithm 1 for several values of $M$ and $r_{min}$. First, we can observe that the energy consumption corresponding to the sequence of solutions is non-increasing and Algorithm 1 converges in few iterations. Moreover, by increasing the number of reflective elements at the IRS $M$, the robot can find paths with lower energy consumptions. The energy consumption increases for higher values of $r_{min}$. This can be better observed in Fig.~\ref{fig:Path2_5}, where we show robot trajectories resulting from Algorithm 1 for $r_{min}=2.5$\, Gbps, and several values of $M$. Specifically, when $r_{min}$ increases, the robot must find paths that are closer to the AP and the IRS to improve the coverage and increase the data rate. Specifically, for all the values of $M$, Algorithm 1 selects MR initial solutions. For $M=0$ and $M=64$ the robot trajectories avoid NLOS areas with respect to the AP, whereas, for $M=128$, paths that are closer to the IRS provide higher rates.

%

\section{Conclusion}
\label{sec:Conc}
In this work, we have proposed a novel robot trajectory optimization problem with QoS constrained communications for minimizing the motion energy consumption. The robot must avoid collisions with obstacles, reach the destination within a deadline, and transmit uplink data to an AP operating at mm-wave frequency bands and assisted by an IRS. We have proposed a solution that accounts for the mutual dependence between the channel conditions and the robot trajectory. Specifically, by exploiting the mm-wave propagation characteristics, we have decoupled the beamforming (at the AP and IRS) and the trajectory optimization problems. The latter is solved by an SCO algorithm for which the convergence to a KKT point is proved. Moreover, given a radio map, the proposed solution can find trajectories that avoid obstacles collisions and adapt to the propagation characteristics of the scenario. Specifically, we have shown how the number of IRS's reflective elements affects the robot trajectory. Future works will investigate the impact of the tasks' deadline on the solution and will consider more scenarios.

\appendices
\section{}
\label{sec:ApA}
To prove Lemma 2, we prove that $\mbox{r}_{k}^{*}$, which is the estimated rate at timeslot $k$, is a convex function of $d_{i,k}$ and $d_{a,k}$. Then, $\bar{\boldsymbol{\mbox{r}}}^{*}$ is convex because it is a sum of convex functions. We first compute the partial derivatives of $\mbox{r}_{k}^{*}$ with respect to $d_{i,k}$ and $d_{a,k}$. These are given by:
\begin{align*} 
\frac{\partial\mbox{r}_{k}^{*}}{\partial d_{i,k}}=\frac{\Big(-\bar{\nu}\bar{A} d_{i,k}^{-\bar{\nu}-1}-\bar{\nu}/2\bar{B}  d_{i,k}^{-\bar{\nu}/2-1} d_{a,k}^{-\bar{\mu}/2}\Big)\frac{p_{t}}{\sigma^{2}}} {F_{k}},\stepcounter{equation}\tag{\theequation} \label{eq:rate_dev1}
\end{align*}
\begin{align*} 
\frac{\partial \mbox{r}_{k}^{*}}{\partial d_{a,k}}=\frac{\Big(-\bar{\mu}\bar{C} d_{a,k}^{-\bar{\mu}-1}-\bar{\mu}/2\bar{B}  d_{i,k}^{-\bar{\nu}/2} d_{a,k}^{-\bar{\mu}/2-1}\Big)\frac{p_{t}}{\sigma^{2}}}{F_{k}},\stepcounter{equation}\tag{\theequation} \label{eq:rate_dev2}
\end{align*}
where, $F_{k}=\mbox{ln}(2)\Bigg(1+\Big(\bar{A} d_{i,k}^{-\bar{\nu}}+ \bar{B}  d_{i,k}^{-\bar{\nu}/2} d_{a,k}^{-\bar{\mu}/2}+ \bar{C} d_{a,k}^{-\bar{\mu}} \Big)\frac{p_{t}}{\sigma^{2}}\Bigg)>0$. Then, the second order partial derivatives are given by:
\begin{align*} 
\frac{\partial^{2} \mbox{r}_{k}^{*}}{\partial d_{i,k}^{2}}&=\frac{\Big(\bar{\nu}(\bar{\nu}+1)\bar{A} d_{i,k}^{-\bar{\nu}-2}+\bar{\nu}/2(\bar{\nu}/2+1)\bar{B}  d_{i,k}^{-\bar{\nu}/2-2} d_{a,k}^{-\bar{\mu}/2}\Big)F_{k}\frac{p_{t}}{\sigma^{2}}}{F_{k}^{2}}\\
&-\frac{\mbox{ln}(2)\Big(-\bar{\nu}\bar{A} d_{i,k}^{-\bar{\nu}-1}-\bar{\nu}/2\bar{B}  d_{i,k}^{-\bar{\nu}/2-1} d_{a,k}^{-\bar{\mu}/2}\Big)^{2}\frac{p_{t}^{2}}{\sigma^{4}}}{F_{k}^{2}},\stepcounter{equation}\tag{\theequation} \label{eq:rate_dev11}
\end{align*}
\begin{align*} 
\frac{\partial^{2} \mbox{r}_{k}^{*}}{\partial d_{a,k}^{2}}&=\frac{\Big(\bar{\mu}(\bar{\mu}+1)\bar{C} d_{a,k}^{-\bar{\mu}-2}+\bar{\mu}/2(\bar{\mu}/2+1)\bar{B}  d_{i,k}^{-\bar{\nu}/2} d_{a,k}^{-\bar{\mu}/2-2}\Big)F_{k}\frac{p_{t}}{\sigma^{2}}}{F_{k}^{2}}\\
&-\frac{\mbox{ln}(2)\Big(-\bar{\mu}\bar{C} d_{a,k}^{-\bar{\mu}-1}-\bar{\mu}/2\bar{B}  d_{i,k}^{-\bar{\nu}/2} d_{a,k}^{-\bar{\mu}/2-1}\Big)^{2}\frac{p_{t}^{2}}{\sigma^{4}}}{F_{k}^{2}},\stepcounter{equation}\tag{\theequation} \label{eq:rate_dev22}
\end{align*}
\begin{align*} 
\frac{\partial^{2} \mbox{r}_{k}^{*}}{\partial d_{i,k} \partial d_{a,k}}&=\frac{\Big((\bar{\mu}/2)(\bar{\nu}/2)\bar{B}  d_{i,k}^{-\bar{\nu}/2-1} d_{a,k}^{-\bar{\mu}/2-1}\Big)F_{k}\frac{p_{t}}{\sigma^{2}}}{F_{k}^{2}}-\\
&\frac{\mbox{ln}(2)\Big(-\bar{\nu}\bar{A} d_{i,k}^{-\bar{\nu}-1}-\bar{\nu}/2\bar{B}  d_{i,k}^{-\bar{\nu}/2-1} d_{a,k}^{-\bar{\mu}/2}\Big)\frac{p_{t}^{2}}{\sigma^{4}}}{F_{k}^{2}} \times\\
&\frac{\Big(-\bar{\mu}\bar{C} d_{a,k}^{-\bar{\mu}-1}-\bar{\mu}/2\bar{B}  d_{i,k}^{-\bar{\nu}/2} d_{a,k}^{-\bar{\mu}/2-1}\Big)\frac{p_{t}^{2}}{\sigma^{4}}}{F_{k}^{2}}.\stepcounter{equation}\tag{\theequation} \label{eq:rate_dev12}
\end{align*}
We can verify that $\frac{\partial^{2} \mbox{r}_{k}^{*}}{\partial d_{i,k}^{2}}>0$, $\frac{\partial^{2} \mbox{r}_{k}^{*}}{\partial d_{a,k}^{2}}>0$ and $\frac{\partial^{2} \mbox{r}_{k}^{*}}{\partial d_{i,k}^{2}}\frac{\partial^{2} \mbox{r}_{k}^{*}}{\partial d_{a,k}^{2}}-\Bigg(\frac{\partial^{2}\mbox{r}_{k}^{*}}{\partial d_{i,k} \partial d_{a,k}}\Bigg)^{2}>0$. Therefore, the Hessian is positive definite and $\mbox{r}_{k}^{*}$ is a convex function of $d_{i,k}$ and $d_{a,k}$. 

\section{}
\label{sec:ApB}
We prove Lemma 3 by following the same reasoning of Appendix A. Note that $\bar{\boldsymbol{\mbox{r}}}_{app}^{*}$ in~\eqref{eq:avg_rate_appx} is the sum of $K+1$ functions each of them depending only on the robot's position at timeslot $k$: 
\begin{align*} 
&\bar{\boldsymbol{\mbox{r}}}_{app}^{*}=\frac{1}{K}\sum_{k=0}^{K}\mbox{r}_{app,k}^{*}=\stepcounter{equation}\tag{\theequation} \label{eq:rappk}\\
&\frac{1}{K}\sum_{k=0}^{K}B_{w}\mbox{log}_{2}\left(1+\Big(\bar{A} d_{i,0,k}^{-\bar{\nu}}+ \bar{B}  d_{i,0,k}^{-\bar{\nu}/2} d_{a,0,k}^{-\bar{\mu}/2}+\bar{C} d_{a,0,k}^{-\bar{\mu}} \Big) \frac{p_{t}}{\sigma^{2}}\right)\\
&-\frac{\partial \mbox{r}_{k}^{*}}{\partial d_{a,k}}|_{\big(d_{a,0,k},d_{i,0,k}\big)} d_{a,0,k}-\frac{\partial \mbox{r}_{k}^{*}}{\partial d_{i,k}}|_{\big(d_{a,0,k},d_{i,0,k}\big)} d_{i,0,k}\\&+\frac{\partial \mbox{r}_{k}^{*}}{\partial d_{a,k}}|_{\big(d_{a,0,k},d_{i,0,k}\big)}\sqrt{(z_{r}-z_{a})^{2}+(x_{k}-x_{a})^{2}+(y_{k}-y_{a})^{2}}\\
&+\frac{\partial \mbox{r}_{k}^{*}}{\partial d_{i,k}}|_{\big(d_{a,0,k},d_{i,0,k}\big)}\sqrt{(z_{r}-z_{i})^{2}+(x_{k}-x_{i})^{2}+(y_{k}-y_{i})^{2}},\\
\end{align*}
where, $\frac{\partial \mbox{r}_{k}^{*}}{\partial d_{i,k}}<0$ and $\frac{\partial \mbox{r}_{k}^{*}}{\partial d_{a,k}}<0$ are given by~\eqref{eq:rate_dev1} and~\eqref{eq:rate_dev2}, respectively. Note that in~\eqref{eq:rappk} we have replaced $q_{k}$ with $[x_{k},y_{k}]$. Let us define $D_{k}=\sqrt{(z_{r}-z_{a})^{2}+(x_{k}-x_{a})^{2}+(y_{k}-y_{a})^{2}}$, then, we can compute the partial derivatives of $\mbox{r}_{app,k}^{*}$ with respect to $x_{k}$ and $y_{k}$ as follows:
\begin{align*} 
\frac{\partial\mbox{r}_{app,k}^{*}}{\partial x_{k}}=&\frac{\partial \mbox{r}_{k}^{*}}{\partial d_{a,k}}|_{\big(d_{a,0,k},d_{i,0,k}\big)}\frac{(x_{k}-x_{a})} {D_{k}}+\\
&\frac{\partial \mbox{r}_{k}^{*}}{\partial d_{i,k}}|_{\big(d_{a,0,k},d_{i,0,k}\big)}\frac{(x_{k}-x_{i})} {D_{k}},\stepcounter{equation}\tag{\theequation} \label{eq:rate_app_dev1}
\end{align*}
\begin{align*} 
\frac{\partial\mbox{r}_{app,k}^{*}}{\partial y_{k}}=&\frac{\partial \mbox{r}_{k}^{*}}{\partial d_{a,k}}|_{\big(d_{a,0,k},d_{i,0,k}\big)}\frac{(y_{k}-y_{a})} {D_{k}}+\\
&\frac{\partial \mbox{r}_{k}^{*}}{\partial d_{i,k}}|_{\big(d_{a,0,k},d_{i,0,k}\big)}\frac{(y_{k}-y_{i})} {D_{k}}.\stepcounter{equation}\tag{\theequation} \label{eq:rate_app_dev2}
\end{align*}
The second order partial derivatives are given by:
\begin{align*} 
\frac{\partial^{2}\mbox{r}_{app,k}^{*}}{\partial x_{k}^{2}}=&\frac{\partial \mbox{r}_{k}^{*}}{\partial d_{a,k}}|_{\big(d_{a,0,k},d_{i,0,k}\big)}\frac{(y_{k}-y_{a})^{2}+(z_{k}-z_{a})^{2}} {D_{k}^{3}}+\\
&\frac{\partial \mbox{r}_{k}^{*}}{\partial d_{i,k}}|_{\big(d_{a,0,k},d_{i,0,k}\big)}\frac{(y_{k}-y_{i})^{2}+(z_{k}-z_{i})^{2}} {D_{k}^{3}},\stepcounter{equation}\tag{\theequation} \label{eq:rate_app_dev11}
\end{align*}
\begin{align*} 
\frac{\partial^{2}\mbox{r}_{app,k}^{*}}{\partial y_{k}^{2}}=&\frac{\partial \mbox{r}_{k}^{*}}{\partial d_{a,k}}|_{\big(d_{a,0,k},d_{i,0,k}\big)}\frac{(x_{k}-x_{a})^{2}+(z_{k}-z_{a})^{2}} {D_{k}^{3}}+\\
&\frac{\partial \mbox{r}_{k}^{*}}{\partial d_{i,k}}|_{\big(d_{a,0,k},d_{i,0,k}\big)}\frac{(x_{k}-x_{i})^{2}+(z_{k}-z_{i})^{2}} {D_{k}^{3}},\stepcounter{equation}\tag{\theequation} \label{eq:rate_app_dev22}
\end{align*}
\begin{align*} 
\frac{\partial^{2}\mbox{r}_{app,k}^{*}}{\partial x_{k} \partial y_{k}}=&\frac{\partial \mbox{r}_{k}^{*}}{\partial d_{a,k}}|_{\big(d_{a,0,k},d_{i,0,k}\big)}\frac{(x_{k}-x_{a})(y_{k}-y_{a})} {D_{k}^{3}}+\\
&\frac{\partial \mbox{r}_{k}^{*}}{\partial d_{i,k}}|_{\big(d_{a,0,k},d_{i,0,k}\big)}\frac{(x_{k}-x_{i})(y_{k}-y_{i})} {D_{k}^{3}}.\stepcounter{equation}\tag{\theequation} \label{eq:rate_app_dev12}
\end{align*}
Note that $\frac{\partial^{2}\mbox{r}_{app,k}^{*}}{\partial x_{k}^{2}}<0$ and $\frac{\partial^{2}\mbox{r}_{app,k}^{*}}{\partial y_{k}^{2}}<0$ because $\frac{\partial \mbox{r}_{k}^{*}}{\partial d_{i,k}}|_{\big(d_{a,0,k},d_{i,0,k}\big)}<0$ and  $\frac{\partial \mbox{r}_{k}^{*}}{\partial d_{a,k}}|_{\big(d_{a,0,k},d_{i,0,k}\big)}<0$ are negative terms $\forall d_{a,0,k}>0, \forall d_{i,0,k}>0 $. Furthermore, we have that $\frac{\partial^{2}\mbox{r}_{app,k}^{*}}{\partial x_{k}^{2}}\frac{\partial^{2}\mbox{r}_{app,k}^{*}}{\partial y_{k}^{2}}-\Bigg(\frac{\partial^{2}\mbox{r}_{app,k}^{*}}{\partial x_{k} \partial y_{k}}\Bigg)^{2}>0$. Therefore, the Hessian is negative definite and $\mbox{r}_{app,k}^{*}$ is a concave function of $q_{k}=[x_{k},y_{k}]$.


\bibliography{ref}

\begin{thebibliography}{10}
\providecommand{\url}[1]{#1}
\csname url@samestyle\endcsname
\providecommand{\newblock}{\relax}
\providecommand{\bibinfo}[2]{#2}
\providecommand{\BIBentrySTDinterwordspacing}{\spaceskip=0pt\relax}
\providecommand{\BIBentryALTinterwordstretchfactor}{4}
\providecommand{\BIBentryALTinterwordspacing}{\spaceskip=\fontdimen2\font plus
\BIBentryALTinterwordstretchfactor\fontdimen3\font minus
  \fontdimen4\font\relax}
\providecommand{\BIBforeignlanguage}[2]{{%
\expandafter\ifx\csname l@#1\endcsname\relax
\typeout{** WARNING: IEEEtran.bst: No hyphenation pattern has been}%
\typeout{** loaded for the language `#1'. Using the pattern for}%
\typeout{** the default language instead.}%
\else
\language=\csname l@#1\endcsname
\fi
#2}}
\providecommand{\BIBdecl}{\relax}
\BIBdecl

\bibitem{Ericsson1}
R.~Sabella, A.~Thuelig, M.~C. Carrozza, and M.~Ippolito, ``Industrial
  automation enabled by robotics, machine intelligence and 5{G},'' Ericsson
  Technology Review, 2018.

\bibitem{6G}
M.~{Giordani}, M.~{Polese}, M.~{Mezzavilla}, S.~{Rangan}, and M.~{Zorzi},
  ``Toward 6{G} networks: Use cases and technologies,'' \emph{IEEE
  Communications Magazine}, vol.~58, no.~3, pp. 55--61, 2020.

\bibitem{Cheffena}
M.~{Cheffena}, ``Industrial wireless communications over the millimeter wave
  spectrum: opportunities and challenges,'' \emph{IEEE Communications
  Magazine}, vol.~54, no.~9, pp. 66--72, Sep. 2016.

\bibitem{IRSmmwave3}
P.~{Wang}, J.~{Fang}, X.~{Yuan}, Z.~{Chen}, and H.~{Li}, ``Intelligent
  reflecting surface-assisted millimeter wave communications: Joint active and
  passive precoding design,'' \emph{IEEE Transactions on Vehicular Technology},
  vol.~69, no.~12, pp. 14\,960--14\,973, 2020.

\bibitem{En1}
{Yongguo Mei}, {Yung-Hsiang Lu}, Y.~C. {Hu}, and C.~S.~G. {Lee},
  ``Energy-efficient motion planning for mobile robots,'' in \emph{IEEE
  International Conference on Robotics and Automation, Proceedings ICRA '04},
  vol.~5, 2004, pp. 4344--4349 Vol.5.

\bibitem{CoCP}
Y.~{Yan} and Y.~{Mostofi}, ``Co-optimization of communication and motion
  planning of a robotic operation under resource constraints and in fading
  environments,'' \emph{IEEE Transactions on Wireless Communications}, vol.~12,
  no.~4, pp. 1562--1572, 2013.

\bibitem{UAVmmWave1}
Z.~{Xiao}, P.~{Xia}, and X.~{Xia}, ``Enabling {UAV} cellular with
  millimeter-wave communication: potentials and approaches,'' \emph{IEEE
  Communications Magazine}, vol.~54, no.~5, pp. 66--73, 2016.

\bibitem{UAVIRS2}
S.~{Li}, B.~{Duo}, X.~{Yuan}, Y.~{Liang}, and M.~{Di Renzo}, ``Reconfigurable
  intelligent surface assisted {UAV} communication: Joint trajectory design and
  passive beamforming,'' \emph{IEEE Wireless Communications Letters}, vol.~9,
  no.~5, pp. 716--720, 2020.

\bibitem{Borrelli}
X.~{Zhang}, A.~{Liniger}, and F.~{Borrelli}, ``Optimization-based collision
  avoidance,'' \emph{IEEE Transactions on Control Systems Technology}, pp.
  1--12, 2020.

\bibitem{Conv}
\BIBentryALTinterwordspacing
B.~R. Marks and G.~P. Wright, ``A general inner approximation algorithm for
  nonconvex mathematical programs,'' \emph{Operations Research}, vol.~26,
  no.~4, pp. 681--683, 1978. [Online]. Available:
  \url{http://www.jstor.org/stable/169728}
\BIBentrySTDinterwordspacing

\bibitem{MAPP}
C.~{Tatino}, N.~{Pappas}, and D.~{Yuan}, ``Multi-robot association-path
  planning in millimeter-wave industrial scenarios,'' \emph{IEEE Networking
  Letters}, vol.~2, no.~4, pp. 190--194, 2020.

\bibitem{MMInd2}
S.~{Saponara}, F.~{Giannetti}, B.~{Neri}, and G.~{Anastasi}, ``Exploiting
  mm-wave communications to boost the performance of industrial wireless
  networks,'' \emph{IEEE Transactions on Industrial Informatics}, vol.~13,
  no.~3, pp. 1460--1470, 2017.

\end{thebibliography}
\bibliographystyle{IEEEtran}

\end{document}